\documentstyle[12pt]{article}

\newcommand{\sect}[1]{\setcounter{equation}{0}\section{#1}}

\def\noi{\noindent}
\def\ar{\begin{array}{rcl}}
\def\an{\end{array}}
\def\ra{\rangle}
\def\la{\langle}
\newcommand{\eq}{\begin{equation}}
\newcommand{\eqa}{\begin{eqnarray}}
\newcommand{\en}{\end{equation}}
\newcommand{\ena}{\end{eqnarray}}
\def\ef{{\scriptscriptstyle {\cal F}_{12}}}

\def\Hil{{\cal H}}
\def\ot{\otimes}
\def\id{\mbox{id}}
\def\ie{\mbox{\it i.e.\/ }}
\def\eg{\mbox{\it e.g.\/ }}

\def\g{\mbox{\bf g}}
\def\uqg{\mbox{$U_{q}{\/\mbox{\bf g}}$}}

\newcommand{\tr}{\triangleright}
\newcommand{\trs}{{\,\stackrel{s}{\triangleright}\,}}

\def\R{\mbox{$\cal R$}}

\def\A{\mbox{$\cal A$}}

\def\F{\mbox{$\cal F$}}
\def\V{\mbox{$\cal V$}}
\def\Op{\mbox{$\cal O$}}

\def\z{\hspace*{9mm}}
\def\x{\hspace{3mm}}
\newcommand{\ad}{\stackrel{\mbox{\scriptsize ad}}{\triangleright}}

\newcommand{\nn}{{\bf N}}

\begin{document}
\begin{titlepage}
\begin{center}
August 1995          \hfill       LMU-TPW 95-10\\
\mbox{}            \hfill       (revised, Feb.\ 1996)\\
\vskip.6in

{\Large \bf Identical Particles and Quantum Symmetries}

\vskip.4in

Gaetano Fiore* and Peter Schupp

\vskip.25in

{\em Sektion Physik der Ludwig-Maximilians-Universit\"at
M\"unchen\\
Theoretische Physik --- Lehrstuhl Professor Wess\\
Theresienstra\ss e 37, 80333 M\"unchen\\
Federal Republic of Germany}
\end{center}
\vskip1in
\begin{abstract}
We propose a solution to the problem of compatibility of
Bose-Fermi statistics
with symmetry transformations implemented by 
compact quantum groups of Drinfel'd type.
We use unitary transformations to conjugate multi-particle
symmetry postulates, so as to obtain a twisted realization
of the symmetric groups $S_n$.
\end{abstract}
\vfill
\noi \hrule
\vskip.2cm
\noi{\footnotesize *A.~v.~Humboldt-fellow\hfill {\it e-mail: }
Fiore, Schupp \ @ \ ls-wess.physik.uni-muenchen.de}
\end{titlepage}
\newpage
\setcounter{page}{1}

\sect{Introduction}

Quantum groups \cite{dr,ji,frt} have received much attention in
recent years as
candidates for generalized symmetry transformations in physics.
Among
other applications, they look promising in relation to generalized
space-time\footnote{These are symmetries of a proposed
non-commutative structure of space-time \cite{wess}.}
and/or internal symmetries in Quantum Field Theory.
One way to
approach QFT consists first 
in finding  a consistent procedure
to implement quantum group transformations in Quantum
Mechanics with a finite number of particles, then 
to pass to QFT through second
quantization. Various models describing systems of one
particle (see e.g. ref. \cite{wess1,wess2,fio,fioeu,wei}) or a finite number of
{\em distinct}
particles consistently transforming under the action of a quantum
group have been constructed so far; as known, the quantum group
coproduct plays a specific role in extending quantum group
transformations from one-particle to multi-particle systems.
In this article we would like to study whether the notions of
identical particles and quantum group  transformations
are compatible in quantum mechanics (in
first quantization).

The setting that we have in mind is a quantum
mechanical system transforming under generalized (symmetry)
transformations realized by some $*$-Hopf algebra $H$ \footnote{The
transformations  may correspond to a symmetry
either in the sense that they leave  the {\it dynamics} of
the particular system under consideration invariant  
(e.g. rotation symmetry
of its Hamiltonian), and therefore are associated to conservation
laws for the latter; or in the sense that they leave
the {\it form} of the physical description of {\it any}
system invariant 
(covariance of the physical description), as it happens e.g. with the
Poincar\'e transformations in Special Relativity.} (in particular, a
$*$-quantum group  \cite{dr}). In order that a system
of $n$ bosons/fermions transforms under the action of $H$ its Hilbert space
of states should carry both a representation of the symmetric group $S_n$
and of $H$. 
In the case that the $H$ is quantum group, one might expect that 
this is impossible.

We recall that in the standard quantum mechanical formalism the 
elements of $S_n$ are realized as ordinary permutation operators.
On the other hand, in the Hopf algebra formalism the action of 
$H$ on a multiparticle
system is defined through the coproduct $\Delta$. Given
a representation $\rho$ of $H$ on a ``one-particle'' Hilbert space $\Hil$, and
considering (for simplicity) the case of two particles,  the
action of $H$ on $\Hil \ot \Hil$ is defined through 
$(\rho\ot\rho)\circ \Delta$. In the case that $H$ is cocommutative
(\eg $H=U(su(2))$), the coproduct takes the form 
$\Delta(X_i)= X_i\ot 1 + 1\ot X_i$ on all the generators $X_i$
(in the case $H=U(su(2))$ this expresses the classical addition law
of angular momentum); therefore the above action preserves
the symmetric  and antisymmetric subspaces
$(\Hil \ot \Hil)_\pm$ defined by $P_{12} (\Hil \ot \Hil)_\pm =
\pm (\Hil \ot \Hil)_\pm$ respectively ($P_{12}$ denotes the permutation
operator). When $H$ is not cocommutative, \eg it is a quantum group,
 $\Delta$ is no more symmetric under the action of $P_{12}$, so that
the above action mixes $(\Hil \ot \Hil)_+$ and $(\Hil \ot \Hil)_-$. 
Therefore, fermions and bosons in the ordinary sense seem impossible,
and it is natural to speculate that in the quantum group  context 
some new (or ``$q$-") statistics
is necessary or even that the notion of identical particles must be 
abandoned. 

Even if $H$ is just a {\em slight\/} deformation of a co-commutative Hopf algebra 
(e.g. an ordinary Lie group)
a new statistics would result into a drastic  discontinuity of
the number of allowed states of the multi-particle system in the limit of
vanishing deformation parameter ($\ln q$ in the $H=\uqg$-case):
in fact, elementary particles cannot be ``almost identical'', they
can only be either identical or different. However,
such a  discontinuity appears physically unacceptable if we think of 
$H$ as a slight modification of some experimentally
well-established symmetry of elementary particle physics.

 A previously suggested ``quick fix'' of the problem is the 
naive symmetrization of coproducts 
---this approach will however
destroy any true quantum symmetry.
 It is also important to realize that
it is not enough to make sense of $\Delta(H), \Delta^2(H),$ {\em etc}. 
The spaces of multi-particle operators have to be larger than that 
to be in one-to-one correspondence with their 
classical (symmetrized) counterparts.  In the $H=U_q(su(2))$ case,
for instance, we would like to construct the q-analog of the
(classically) symmetric operators $X_i\ot X_i$, which are not
the coproduct of anything.

In this work we want to show that a solution to the problem is a
modification of  our notions of
symmetry and
anti-symmetry associated to
bosons and fermions. The point is that ordinary permutations
are not the only possible realization of elements of the abstract
group $S_n$; an alternative one can be obtained by applying
some unitary transformation $F_{12\ldots n}$ to the permutators
(see section \ref{twistmult}). The question (see section \ref{qsym}) 
is therefore whether 
for any number of particles
$n$ there exists some $F_{12\ldots n}$ (the ``twist")
such that the corresponding realization 
of $S_n$ is compatible with the action of $H$.
Due to some theorems by Drinfel'd,
this turns out to be the case {\em at least\/} if $H = \uqg$  \cite{dr,ji,frt} is
one of the standard quantum groups associated to the 
compact\footnote{For U${}_q$({\bf g}) this requires $q \in {\bf R}$. 
To study the problem in the case of $q$ on the
unit circle the reader should consult \cite{mascho} for the structure of weak
quasi-Hopf algebras.} 
simple Lie algebras
$\g$ of the classical series---the case of  $U_q \mbox{su}(2)$
will be studied in some detail in section \ref{example}---or 
if $H$ is a triangular Hopf algebra  arising from the
quantization of
a solution of the classical Yang-Baxter equation 
\cite{drinf,Taktadjan}
or from a twist of type \cite{resh} as {\em e.g.}\/ studied
in \cite{luk}\footnote{Twisted coproducts are here interpreted as
clustered 2-particle states.}.
The precise criterion is that $H$ must be the twist of a
co-commutative
(quasi-)Hopf algebra \cite{Drinfeld}; in either case we
also need the existence of a $*$-conjuga\-tion.

In the case where $H$ is a quasi-triangular Hopf algebra one
might have
expected to see anyons arise as a consequence of the
braid group character
of $\R$; however, in our formulation this does not happen: The 
statistics parameter is not modified---bosons stay 
bosons\footnote{See \cite{voza,nelson} and the extended list of 
references therein for a discussion
of this point in the context of $q$-deformed oscillators.}, fermions
stay fermions, and anyons (though not studied explicitly) stay anyons.
The extreme case of $q = -1$ is especially instructive in this
context \cite{zachos,zhang}.

Let us ask now how in the context of identical particles 
the existence of quantum group symmetries 
of the above kind could 
manifest itself experimentally:
The dynamical evolution of a system of  $n$ identical particles will 
contain new physics only if we adopt an Hamiltonian which is {\em natural\/} 
to the twisted picture. One can always obtain a Hamiltonian {\em consistent\/}
with twisted symmetrization postulates by a
unitary transformation (through  $F_{12\ldots n}$) on a Hamiltonian 
corresponding to some undeformed model, however, 
such a Hamiltonian will in general be
of a very complicated, {\em i.e.\/} unnatural form.
In section \ref{onemany} we will analyze a scattering
experiment to see how the  twist  will 
manifest itself in the 
transformation of the initial and final data (which is essentially
the tensor product of one-particle states) into the 
equivalent twisted (anti-)symmetrized states upon which
the evolution operator describing the scattering acts.
The  $F_{12\ldots n}$ can again be absorbed in a redefined Hamiltonian, so
that an
experiment cannot decide whether we are in the twisted picture or not. (It is
just a change of base.) We can only tell what picture is more natural.
The main message is then that the twisted picture can be consistently
introduced. (Contrary to expectation, there are no problems with statistics.)
The twisted picture may lead to the development of models that one would
probably not think of otherwise.
One would expect to see {\em direct\/} consequences of the twists only
in particle
creation and annihilation processes; this however belongs to the realm of
quantum field theory and will be treated elsewhere.

For readers not  familiar with 
the notion of Hopf algebras, we give a very brief  introduction 
to the subject  in section \ref{appendix}.

After completion of this work we became aware of the very interesting
paper in Ref. \cite{wopu}, which gives a quantization scheme for fields
transforming covariantly under $SU_q(N)$. 
In a future work we will compare our results
with the ones therein while considering the issue of second 
quantization.

\sect{Twisted Multi-Particle Description}
\label{twistmult}

Let us forget  the issue of quantum symmetry and
hence the coproduct
for the moment,
and just consider pure quantum mechanics
for identical particles. 
Consider a one-particle
system, denote by $\Hil$ the  Hilbert space of its states,
and by $\A$
the $*$-algebra of observables acting on $\Hil$.
$n$-particle states and $n$-particle operators
will live in as yet to be determined subspaces of $\Hil^{\ot n}$ and
$\A^{\ot n}$ respectively.

Let us consider states of two identical particles. The
corresponding state
vector $|\psi^{(2)}\ra$ will be some element of the tensor product
of two copies of 
the one-particle Hilbert
space $\Hil$. Let $ P_{12}$ be
the permutation operator on
$\Hil \ot \Hil$:\ $ P_{12}(|a\ra \ot |b\ra) \equiv
|b\ra \ot |a\ra$. (In the sequel we will also use the symbol $\tau$ to
denote the abstract
permutation map of two tensor factors, $ \tau(a \ot b) \equiv b \ot a$.)
The fact that we are dealing with identical particles manifests
itself in the properties of state vectors under permutation:
\eq
 P_{12}|\psi^{(2)}\ra = e^{i \nu}|\psi^{(2)}\ra,\label{stat}
\en
where $\nu = 0$ for Bose-statistics and
$\nu = \pi$ for Fermi-statistics. For the corresponding
expectation value of an arbitrary
operator ${\cal O} \in \A \ot \A$ we then find
\eq
\la\psi^{(2)}|{\cal O}|\psi^{(2)}\ra = \la\psi^{(2)}| P_{12}\:
{\cal O}\: P_{12}|\psi^{(2)}\ra
\en
because the phases $e^{-i \nu}$ and $e^{i \nu}$ from the bra and
the ket
cancel. This means that the operators $\cal O$ and
$ \tau({\cal O}) \equiv
 P_{12} {\cal O} P_{12}$ are members of the same
equivalence
class as far as expectation values go.
One particular representative of each  such equivalence
class is the symmetrized operator
\eq
{1 \over 2}({\cal O} +  \tau({\cal O}) ) \in (\A \ot \A)_+.
\en
It plays a special role because it preserves the two-particle Hilbert
spaces
for any statistic (\ref{stat}), as we will recall below.
We can hence avoid redundant operators
by restricting  $\A \ot \A$ to the sub-algebra
 \eq
(\A \ot \A)_+ := \{ a \in \A \ot \A : [P_{12}, a ] = 0  \}
\label{symop}
\en
(note that $[P_{12}, a ] = 0 \Leftrightarrow  \tau(a) = a$).
In this article we will show how to find an analog of $(\A \ot \A)_+$
compatible with quantum group transformations.

We summarize the relevant equations characterizing a system of two
bosons or fermions:
\eqa
&& P_{12}|u\ra_\pm = \pm|u\ra_\pm \z \mbox{for} \x  |u\ra_{\pm} \in
(\Hil \ot \Hil)_\pm \label{fb}\label{from}\\
&& a:(\Hil \ot \Hil)_\pm \rightarrow (\Hil \ot \Hil)_\pm \z \mbox{for}
\x a
\in (\A \ot \A)_+ \label{ahh}\\
&& *_2: (\A \ot \A)_+\rightarrow (\A \ot \A)_+, 
\z\mbox{where}\x *_2 \equiv * \ot * .\label{to}
\ena
Equation (\ref{fb}) defines bosonic $(+)$ and fermionic $(-)$
states as in
(\ref{stat}).
Equation (\ref{ahh}) follows from
$[P_{12}, (\A\ot\A)_+] = 0$
and shows that symmetrized operators transform boson states
into
bosons states and fermion states into fermion states.

Similar statements as given here for two particles
obviously apply also to states of 3 and more identical particles
and to
other statistics (anyons).

Can one also describe in a non-standard way the system of $n$
identical particles,
using what we know for one particle, so that the description is
perfectly
consistent from the physical viewpoint? Let us concentrate on
two-particle
systems for the moment:

For a unitary and in general not symmetric
operator $F_{12} \in \A \ot \A$, $F_{12}^{*_2} = F_{12}^{-1}$
where $*_2 = * \ot *$, we define
\eqa
(\Hil \ot \Hil)_{\pm}^{{F_{12}}} & := & F_{12} (\Hil \ot \Hil)_\pm\\
P_{12}^{F_{12}}& := & F_{12} P_{12} F_{12}^{-1} \\
(\A \ot \A)_+^{F_{12}}& := & F_{12}(\A \ot \A)_+ F_{12}^{-1}
\ena
where $(\A \ot \A)_+$ is as given above.
We then find in complete analogy to equations (\ref{from} --
\ref{to})
\eqa
&& P_{12}^{F_{12}}|u\ra_\pm = \pm|u\ra_\pm \z \mbox{for} \x
|u\ra_\pm \in
(\Hil \ot\Hil)_\pm^{F_{12}}\\
&& a:(\Hil \ot \Hil)^{F_{12}}_\pm \rightarrow (\Hil \ot \Hil)^
{F_{12}}_\pm \z
\mbox{for}\x a \in (\A \ot
\A)_+^{F_{12}}\label{aHH}\\
&& *_2: (\A \ot \A)_+^{F_{12}}\rightarrow (\A \ot \A)_+^{F_{12}}
\ena
and $a^{F_{12}}:= F_{12} a F_{12}^{-1}$ is hermitean iff $a$ is.
Equation (\ref{aHH}) follows from
\eq
[P_{12}^{F_{12}}, (\A\ot\A)_+^{F_{12}}] = 0.
\label{commop}
\en
In general, $(\Hil \ot \Hil)^{F_{12}}_\pm$ will not be
(anti-)symmetric, nor
will $(\A \ot \A)_+^{F_{12}}$ be symmetric.
Can we still interpret $(\Hil \ot \Hil)_\pm^{F_{12}}$ as 
the Hilbert space of states of
the system of two bosons or fermions of equal type and
$(\A \ot \A)_+^{F_{12}}$ as
the corresponding $*$-algebra of observables?
We can. In fact, we have just conjugated the standard description
of the
2-particle system through $F_{12}$ into a unitary equivalent one.
(This agrees with the general viewpoint put forward in \cite{eng,engnote}.
See also next section for a discussion from the physical point of view.)

Obviously the idea of conjugation can  be generalized
to a system of $n$ identical particles:
Let $F_{12\ldots n} \in \A^{\ot n}$
be unitary,
\ie $(F_{12\ldots n})^{*_n} = (F_{12\ldots n})^{-1}$, where $*_n :=
*^{\ot n}$, and define
\eqa
(\Hil \ot\ldots\ot \Hil)_\pm^{F_{12\ldots n}}& := & F_{12\ldots n}
(\Hil \ot\ldots\ot \Hil)_\pm \label{def1}\\
P_{12}^{F_{12\ldots n}}& := & F_{12\ldots n} P_{12}
(F_{12\ldots n})^{-1}\\
& \vdots & \nonumber \\
P_{n-1,n}^{F_{12\ldots n}}& := & F_{12\ldots n} P_{n-1,n}
(F_{12\ldots n})^{-1}\\
(\A \ot\ldots\ot\A)_+^{F_{12\ldots n}}& := & F_{12\ldots n}
(\A\ot\ldots\ot\A)_+(F_{12\ldots n})^{-1}
\label{twistop}
\ena
where
$$
(\A\ot\ldots\ot\A)_+ := \{ a \in \A\ot\ldots\ot\A : [P_{i,i+1} , a] =0 ,
i=1,\ldots n-1\},
$$
and $P_{i,i+1}$ is the permutator of the $i^{th},(i\!+\!1)^{th}$
tensor factors.
Then
\eqa
&&P_{i,i+1}^{F_{12\ldots n}}|u\ra_\pm =
\pm |u\ra_\pm \z\mbox{for}\x |u\ra_\pm \in (\Hil \ot\ldots\ot
\Hil)_\pm^{F_{12\ldots n}} \label{astate}\\
&&a: (\Hil \ot\ldots\ot \Hil)_\pm^{F_{12\ldots n}}\rightarrow
(\Hil \ot\ldots\ot \Hil)_\pm^{F_{12\ldots n}} \label{aHdH}\\
&& \z\z\z\z\z\mbox{for}\x a \in
(\A \ot\ldots\ot\A)_+^{F_{12\ldots n}}\\
&&*_n: (\A \ot \ldots \ot\A)_+^{F_{12\ldots n}}\rightarrow
(\A \ot \ldots \ot\A)_
+^{F_{12\ldots n}}.
\ena
Equation (\ref{aHdH}) follows from
\eq
[P_{i,i+1}^{F_{12\ldots n}}, (\A\ot\ldots\ot\A)_+^
{F_{12\ldots n}}] = 0.
\label{asop}
\en
 Note that  in eqs. (\ref{astate}) to (\ref{asop}) the twist
$F_{12\ldots n}$
does not explicitly appear any more; these equations give an
{\it intrinsic}
characterization of the twisted multi-particle description,
involving only
the operators $P_{i,i+1}^{F_{12\ldots n}}$. 

By construction  $P_{i,i+1}^{F_{12\ldots n}}$
is hermitean, its square is
the identity and (consequently) has only eigenvalues $\pm 1$;
moreover, the degeneracy of these eigenvalues  is the same as
in the untwisted case. The operators $P_{i,i+1}^{F_{12\ldots n}}$
give a realization of the group $S_n$ of permutation of $n$
objects, 
because they satisfy the same algebraic relations as the
ordinary permutators $P_{i,i+1}$; correspondingly, 
$(\A\ot\ldots\ot\A)_+^{F_{12\ldots n}}$, 
$(\Hil \ot\ldots\ot\Hil)_\pm^{F_{12\ldots n}}$ carries
irreducible representations of $S_n$.
Viceversa, one could easily prove that, given
operators satisfying these conditions, one can find a unitary
$F_{12\ldots n}$ such that equations (\ref{def1}) to (\ref{twistop})
hold.

It will turn out that, even though the twists which are  relevant for
the quantum
symmetry issue are very hard to compute, the
$P_{i,i+1}^{F_{12\ldots n}}$
are much less so; see section~\ref{example}.

\noi {\em Remark:} If we replace the nilpotent $P_{12}$ by some
braid group
generator one could also conjugacy transform anyons.

\sect{Identical Versus Distinct Particles }
\label{onemany}

In some situations particles of the same kind can be equivalently
treated as {\it identical} or {\it distinct}, and
there exists a precise correspondence between these two
descriptions.
The twist $F$ directly enters the rule governing this 
correspondence
while in the twisted postulates (\ref{astate}) to (\ref{asop})
(intrinsic formulation) it
appears only hidden in the $P^F$ (together with its inverse).
Transforming one kind of description into the other one
is often needed for practical purposes, as we illustrate
by the following example.

Consider a gedanken experiment of a scattering of
two identical particles. One can distinguish three stages.
In the initial stage, the two particles are far
apart and are assumed to be prepared 
in two separate one-particle normalized states $|\psi_1\ra$,
$|\psi_1\ra$ with vanishing overlap.
In the intermediate stage, the particles approach each other and 
scatter. In the
final stage, long after the collision, the particles are
again far apart and are detected by one-particle detectors.
In the initial and final stage we perform essentially
one-particle preparations/measurements, i.e. we have the choice to
treat the
particles as distinct, whereas in the intermediate stage
the collision is correctly described only if we apply a
symmetric evolution operator to a properly
(anti-)symmetrized two-particle state, that is, if we treat
the two particles as identical. The existence
of two equivalent descriptions (``distinct" versus ``identical")
of the two-particle system in the initial and final stages
and the {\it correspondence rule} that relates the two is an essential
ingredient of the standard quantum-mechanical
formalism.
   
In this section we want to determine how
the conditions for the existence of two
equivalent descriptions (``distinct" and ``identical")
and the correspondence rule between the latter are modified
in the twisted formalism.
As a by-product,
we will realize that closed systems can still be described consistently:
If we e.g. want to describe a system of identical particles in our lab 
we are essentially 
allowed to forget about the existence of other particles
of the same kind in the universe.

Let us consider two-particle scattering again:
Let initial states 
$|\psi_1\ra,|\psi_2\ra$ range on some orthogonal
subspaces $\Hil_1, \Hil_2$
of the whole Hilbert space.\footnote{More
generally, if the preparation were uncomplete, we would assume
that particle 1,2 is in a {\it mixture} of states of $\Hil_1, \Hil_2$
respectively.}
\begin{enumerate}
\item[\bf(1)] We can treat the two particles
as {\it distinct} particles described by the state
\eq
|\psi_d\ra:= |\psi_1\ra\ot|\psi_2\ra \z\in\z \Hil_1\ot \Hil_2.
\label{stated}
\en
A measurement  process is 
described
via a two-particle observable ${\cal O}_1\ot {\cal O}_2$,
${\cal O}_i:\Hil_i\rightarrow \Hil_i$; the probability amplitude
to find  the two-particle system in a state
$|\psi_d'\ra:= |\psi_1'\ra\ot|\psi_2'\ra$
is 
$\la\psi_d|\psi_d'\ra=\la\psi_2|\psi_2'\ra\la\psi_1|\psi_1'\ra$.
This amounts respectively to measuring
${\cal O}_1$ on the first
{\it and} ${\cal O}_2$ on the second, and to the probability
amplitude to find particle 1 in state $|\psi_1'\ra$
{\it and} particle 2 in state $|\psi_2'\ra$.
In particular, setting ${\cal O}_2=id$, $|\psi_2'\ra=|\psi_2\ra$
means that we neglect the information that we have on the second particle, 
\ie we ignore its existence.
\item[\bf(2)] We can treat them as {\it identical}  particles
forming a two-particle system
and describe the latter by the twisted (anti)symmetrized state
\eq
|\psi\ra=
P^{F_{12}}_{S/A}|\psi_d\ra:=\frac {F_{12}}{\sqrt{2}}\left(
|\psi_1\ra\ot|\psi_2\ra
\pm |\psi_2\ra
\ot|\psi_1\ra\right)\in (\Hil\ot\Hil)^{F_{12}}_{\pm}
\label{sstate}
\en
of bosons $(+)$ or fermions $(-)$.
The measurement
process of (1) is now described by acting on
$|\psi\ra$
through the twisted symmetrized two-particle observable
$F_{12}\left({\cal O}_1\ot {\cal O}_2+{\cal O}_2\ot {\cal
O}_1\right)F_{12}^{-1}\in (\A\ot A)^{F_{12}}_+$.
\end{enumerate}
Description (2) is perfectly equivalent to (1) because 
the mapping (1)$\rightarrow$(2) preserves
scalar products between states (\ie probability 
amplitudes: 
$\la\psi'|\psi\ra=\la\psi_d'|\psi_d\ra=
\la\psi_1'|\psi_1\ra\la\psi_2'|\psi_2\ra)$
and spectra of the observables (\ie results of
measurements).

For the dynamical evolution, including the collision, it
is necessary to use description (2), which involves in an
essential way the quantum statistics.
Nevertheless, if for later
times the state $|\psi\ra(t)$ becomes a combination of states
of the form
\eq
|\psi\ra(t)=\sum\limits_{i,j}a_{ij}
\frac {F_{12}}{\sqrt{2}}\left(
|i\ra\ot|j\ra\pm |j\ra\ot|i\ra\right),
\en
where $|i\ra\in {\cal H}_1'$, $|j\ra\in {\cal H}_2'$, and
${\cal H}_1'$, ${\cal H}_2'$ are {\it orthogonal} subspaces
of ${\cal H}$ (describing e.g. the states of the particle
in detectors 1,2 respectively) then
description (1) can be implemented again:
we can apply $F^{-1}_{12}$ and drop the 
(anti-)symmetrization to get the state
\eq
|\psi_d\ra(t)=\sum\limits_{i,j}a_{ij}
|i\ra\ot|j\ra,
\en
which will give the final correlation between the
potential measurements in the two detectors.

The case of more than two particles can be treated in analogy to the 
case of two particles. Now however we will want to split the 
collection of particles into two (or more) {\em subsystems\/} instead 
of into single particles. If there is negligeable overlap between 
subsystems we are again not forced to treat {\em all\/} particles as
identical particles; we can describe 
particles belonging to different subsystems as distinct, but 
we still have to twist (anti-)symmetrize 
each subsystem.

If we look at the dynamical evolution,
then the same considerations as in the case of two particles will
apply. In particular
as long as the interaction (of any kind) between a 
subsystems and the remaining particles is
negligeable then we have the choice to consider one  subsystem
as isolated (implying that we forget the other particles) or
of treating all particles
as identical.

 These considerations hold
also when the total number of particles of one kind is very large (virtually infinite)
compared to the number in one subsystem. Take this subsystem to 
be our laboratory and we see that
as in the standard formulation, 
to compute any concrete prediction
we can but we don't have  to consider all particles of the given
type present in the universe at the same time [description ``identical''],
namely we may ignore the  ones ``outside our laboratory''
[description ``distinct'']. In principle however we could
apply the postulates of identical particles, through description 
``identical'',
to {\it all} particles of the same type in the universe, without
finding
inconsistent predictions. In other words, the twisted postulates of
Quantum Mechanics for identical particles are completely general and
self-consistent.

\sect{Quantum Symmetries}
\label{qsym}

While their introduction was shown to be consistent,
there was so far no need for the $F_{12\ldots n}$. Now we take the
issue of
quantum group symmetries into consideration.

The picture we have in mind is that of a multi-particle
quantum mechanical model (consisting of identical particles)
on which we  would like to implement generalized (symmetry)
transformations through the action of a generic
Hopf algebra $H$.\footnote{Later we will concentrate on the case of a
twisted image of a cocommutative (quasi-) Hopf algebra;
{\em e.g.\/} $U_q(g)$.}
As given data we take the constituent  one-particle
system, governed by a $*$-algebra $\A$ of
operators that act on a Hilbert space $\Hil$, a $*$-Hopf algebra
$H$ with coproduct $\Delta$, counit $\varepsilon$, antipode $S$ and
complex conjugation $*$, and a unitary realization $\rho$ of $H$ in $\A$.

The key idea that leads to a construction of multi-particle systems that
consistently transform under Hopf algebra actions is that properties of the
coproduct should have to do with
(twisted) (anti-)symmetry of states and operators. 
We will find that coproducts should
be considered as being (twisted) symmetric---even when we are
dealing with
non-cocommutative Hopf algebras as symmetries.

Let us start by recalling what it means that a one-particle system
transforms
under the action of $H$.

\subsection{One-Particle Transformations}
\label{onep}

To begin, we need a representation $\rho$ of $H$ on $\Hil$ which
realizes $H$ in $\A$:\footnote{A given algebra of operators might
first have
to be extended for this scope.}
\eq
\rho\, : \: H \: \rightarrow \: \A;
\en
the map $\rho$ is linear and an algebra homomorphism
$\rho(x y) = \rho(x)\rho(y)$; $\rho(1_H) = 1_{\cal A}$ 
is the identity operator on
$\Hil$.
It is called a unitary representation if in addition
\eq
\rho(x)^* = \rho(x^*).
\label{unitrep}
\en
(For a representation that is not unitary we would find
in contrast
$\rho(x)^* = \overline{\rho^\vee}( x^*)$,
where $\overline{\rho^\vee}$ is the complex conjugate of the
contragredient
representation. For a matrix representation:
$(T^\vee)^i{}_j = S(T^j{}_i) = (T^{-1})^j{}_i$. )

Let $x \in H$, $\Op \in \A$ and $|\psi\ra \in \Hil$.
The actions of $x$ on the one-particle states $|\psi\ra$ and
and $\Op |\psi\ra$ are given
via $\rho$
\eqa
x \tr |\psi\ra & = & \rho(x) |\psi\ra,\label{actsta1}\\
x \tr \big(\Op |\psi\ra \big) & = & \rho(x) \Op |\psi\ra,
\label{actsta2}
\ena
while on the other hand the action of $x$ on the product
$\Op |\psi\ra$ (that is, on an element
of the bigger $H$-module containing both $\A$ and $\Hil$)
should be computed with the coproduct $\Delta$, \ie
\eq
x \tr \big(\Op |\psi\ra \big) = (x_{(1)} \trs \Op) (x_{(2)} \tr |\psi\ra ).
\label{actsta3}
\en
Here and in the sequel we will use Sweedler's notation
$\Delta(x)\equiv x_{(1)}\ot x_{(2)} $ for the coproduct (in the RHS
a sum 
$\sum_i x^i_{(1)}\ot x^i_{(2)}$
of many terms is implicitly understood); similarly,
$\Delta^{(n-1)}(x) \equiv x_{(1)}\ot\ldots\ot x_{(n)}$
for the $(n\!-\!1)$-fold coproduct in Sweedler's notation.
As known, it follows that the action of $H$ on the one-particle
operator $\Op$ is given by\footnote{See however the remark on page~\ref{rem}.}
\eq
x \trs \Op = \rho(x_{(1)}) \:\Op\: \rho(S x_{(2)}),
\qquad x \in H,\ \Op \in \A.
\label{adjoint}
\en
As a concrete example, the reader may think of the
case of quantum mechanics in ordinary three-dimensional space
with transformations consisting of ordinary rotations; in that case
$H$ is the (undeformed) universal enveloping algebra
$U(su(2))$ of the
(covering of the) Lie group $SO(3)$. $\rho$ maps elements of
$U (su(2))$
into operators acting on $\Hil$, out of which we can single out
unitary
operators ``$U$'' realizing finite rotations (i.e. elements of $SO(3)$),
as well as
hermitean ones ``$x$'' realizing infinitesimal rotations (i.e. elements
of $su(2)$)
and generating the whole algebra; in these two cases the action
(\ref{adjoint}) reduces respectively
to conjugation $U\Op U^{-1}$ and to taking the
commutator
$[ix,\Op]$. A rotation
symmetry of the Hamiltonian usually turns elements of
$\rho(U(su(2))$
(\eg angular momentum components) into useful observables for
studying the dynamics of the system.

\subsubsection{Unitary Transformations}

Hermitean conjugation turns an element of $\Hil$, a ``ket'',
into
a ``bra'' which lives in $\Hil^*$ and transforms under the
contragredient
representation. This picture should be preserved under
transformations.
As we know, in the classical case only unitary and---in the
infinitesimal case---anti-hermitean transformation
operators have the required property. In the general Hopf
algebra case
the required property is
$S(x) = x^*$; we will call such elements of $H$
{\em quantum unitary}.
We stress the point that there are two notions
of unitarity which should not be confused: that of a
representation,
and that of a transformation.
Quantum unitary elements also leave
the $*$-structure of $\A$ invariant \cite{thesis}. The condition for
an element $u \in H$ to satisfy
\eq
(u \trs \Op)^* = u \trs \Op^* \qquad \forall \Op \in A \label{requ}
\en
is again
\eq
u^* = S(u) \qquad \label{quop}\mbox{(quantum unitary operator)}.
\en
This is seen as follows:
$*$-conjugating both sides of equation
(\ref{adjoint}) we find a condition
\eq
\rho(S u_{(2)})^* \ot \rho(u_{(1)})^* \stackrel{!}{=}
\rho(u_{(1)}) \ot \rho(S u_{(2)}),
\en
or, using that $\rho$ is a unitary representation,
\eq
(S u_{(2)})^*\ot (u_{(1)})^* \stackrel{!}{=} u_{(1)}\ot S u_{(2)}.
\en
Taking the counit $(\varepsilon \ot id)$ of this equation gives
condition (\ref{quop}). A straightforward calculation that uses 
again unitarity
of the representation $\rho$ and standard facts about $*$-Hopf algebras, like \
$* \circ S = S^{-1} \circ *$ shows that condition (\ref{quop}) is in fact
sufficient for (\ref{requ}).

{\em Remark:}\/ There exist pathological Hopf algebras
(\eg with $\tau\circ\Delta = (id \ot S^2)\Delta$) that
are not $*$-Hopf algebras but still
allow  unitary transformations in a non-standard way.

\subsection{Multi-Particle Transformations}

To implement symmetry transformations (the action of $H$)
on multi-particle systems one makes use of the
coproduct of $H$,
which enters the game in two essentially  different ways.

First, the coproduct is needed to extend the action of $H$ from
one-particle {\it states}  to  $n$-particle states in a way that
preserves the twisted (anti)-symmetry of identical particle
states.
This will constrain  the choice of $F$
in section~\ref{twistmult}, and consequently also the twisted
symmetry of
operators, according to formula (\ref{twistop}). On the other hand,
the coproduct also enters the action of $H$ on single and multiparticle
operators $\Op^{(n)}$ [see formula (\ref{adjoint}) for the
one-particle case];
if the particles are identical this action should again preserve the
twisted symmetry of the operators.
It turns out that both consistency requirements can be simultaneously
satisfied through
an appropriate choice of the $F$'s.

\subsubsection{Transformation of States}
\label{transta}

We have so far required that $\Hil$ be a $*$ $H$-module, \ie that
it carries
a $*$ representation of $H$.
The main task in constructing Hilbert spaces for identical particles
is then to find an operation of twist (anti-)
sym\-me\-tri\-za\-tion that is
compatible with the action of $H$, \ie compatible with the quantum
symmetry transformations.
The action of $H$ on a multi-particle Hilbert space is given once
$\rho^{(n)}$ is known. A representation $\rho$ on the
1-particle Hilbert space
extends to a unitary representation on the $n$-particle Hilbert space
via the $(n-1)$-fold coproduct of $H$:
\eq
\rho^{(n)} = \rho^{\ot n} \circ \Delta^{(n-1)}\,: \:
H \: \rightarrow \: \A^{\ot n} \,:
\Hil^{\ot n} \: \rightarrow \Hil^{\ot n}.
\en
If $\rho$ is unitary then so is $\rho^{(n)}$,
$\rho^{(n)}(x)^{*_n} = \rho^{(n)}(x^*)$, because
$(* \ot *) \circ \Delta= \Delta \circ *$.

Let $x \in H$ and $|\psi^{(n)}\ra \in \Hil^{\ot n}$, then
\eq
x \tr |\psi^{(n)}\ra = \rho^{(n)}(x)|\psi^{(n)}\ra =
\rho(x_{(1)}) \ot\ldots\ot \rho(x_{(n)})|\psi^{(n)}\ra.
\label{actstam}
\en
As always we will first consider the case of two particles.
Similar considerations
will apply to the
case of $n\ge 3$ particles. 
Let $P_{12}$ be the permutation operator on $\Hil \ot \Hil$. 
\paragraph{Symmetric coproduct} 
In the
case of a {\em co-commutative}\/ (\ie symmetric under permutation)
coproduct we have
$$
P_{12}\: \Big((\rho \ot \rho)\Delta_c(x)\Big)
= \Big((\rho \ot \rho)\Delta_c(x)\Big)\: P_{12}
$$
and hence
$$
P_{12} (x \tr |\psi^{(2)}\ra) = x \tr (P_{12}|\psi^{(2)}\ra).
$$
This fact allows us to define symmetrizers
$P_S = {1 \over 2}(I + P_{12})$
and anti-symmetrizers $P_A = {1 \over 2}(I - P_{12})$ that
commute with
the action of $x$, and (anti-) symmetrized Hilbert spaces
\eqa
P_S(\Hil \ot \Hil) & \equiv & (\Hil \ot \Hil)_+,\\
P_A(\Hil \ot \Hil) & \equiv & (\Hil \ot \Hil)_-,
\ena
that are invariant under the action of $x$.  This happens
for instance if $H = U(\g)$, $\g = ${\it Lie}$(G)$. Then $U(\g)$ is
generated
by primitive elements $X_i$ with coproduct
\eq
\Delta^{(n)}(X_i) = \Delta_c^{(n)}(X_i) = X_i \ot 1 \ot\ldots\ot 1 +
\ldots + 1 \ot\ldots\ot X_i;
\en
$\Delta_c^{(n)}(X_i)$ is invariant under permutations and we can
set
$F_{12\ldots n} ={\bf 1}\ot\ldots\ot {\bf 1}$.
\paragraph{Deformed coproduct}
If the coproduct is
{\em not co-commutative}, as it happens for a generic Hopf
algebra,
then the problem arises that the action of $H$ on $(\Hil \ot \Hil)$
will no more preserve the subspaces $(\Hil \ot \Hil)_{\pm}$.
While we should not change the form
of the coproduct (it is at the very heart of quantum groups and
tells us how to act on tensor products) we may however modify our
notion of symmetric operators and (anti-) symmetrized Hilbert
spaces. We can require that
\eq
\rho^{(n)}(H) \subset (\underbrace{\A
\ot\ldots\ot\A}_{\mbox{$n$-times}})_+^{F_{12\ldots n}}
\label{delop}
\en
for some $F_{12\ldots n}$,
so that the
system of $n$ identical particles carries a $*$-representation of
$H$ as well.
This is certainly satisfied if
\eq
\rho^{(n)}(X) = F_{12..n} \rho^{(n)}_c(X) F_{12..n}^{-1},
 \label{deltwist}
\en
where $\rho^{(n)}_c:=\rho^{\ot n} \circ \Delta^{(n-1)}$ and
$\Delta_c$ is a co-commutative coproduct. Equation (\ref{deltwist}) 
has to be read as
a condition on both $\Delta_c$ and $F_{12..n}$.

If $H = \uqg$ \cite{dr,ji,frt},  where $\g$ is the Lie algebra of one of
the simple
Lie groups of the classical series, the following theorem due to
Drinfel'd and Kohno will be our
guidance to the correct choice of the $F$'s we need to satisfy
equations
(\ref{delop}) and (\ref{deltwist}):

\noi
{\bf Drinfel'd Proposition 3.16 in Ref. \cite{Drinfeld}}

\begin{enumerate}
{\em
\item There exists an algebra isomorphism
$\phi: \uqg \stackrel{~}{\leftrightarrow}
(U \g)([[h]])$, where $h = \ln q$ is the deformation parameter.
\item If we identify the isomorphic elements of $\uqg$ and
$(U \g)([[h]])$
then there exists an $\F \in \uqg \ot \uqg$ such that:
\eq
\Delta(a) = \F \Delta_c(a) \F^{-1},\z \forall a \in \uqg \stackrel{~}{=}
(U \g)([[h]]) \label{fdf}
\en
where $\Delta$ is the coproduct of $\uqg$ and $\Delta_c$ is the
(co-commutative) coproduct of $U(\g)$.
\item $(U \g)([[h]])$ is a quasi-triangular quasi-Hopf algebra
(QTQHA)
with universal $\R_\Phi = q^{t/2}$ and a
quasi-coassociative structure given by an element $\Phi \in
\left((U \g)^{\ot 3}([[h]])\right)$ that is expressible in terms of $\F$.
$(U \g)([[h]])$ as QTQHA can be transformed via the twist by $\F$
into the
quasi-triangular Hopf algebra $\uqg$; in particular, the universal
$\R$
of $\uqg$ is given by $\R= \F_{21}\R_\Phi \F^{-1}$. }
\end{enumerate}
Here $(U \g)([[h]])$ denotes the algebra of formal power series in
the
elements of a  basis of $\g$, with coefficients being
entire functions of $h$; $(U \g)([[h]])|_{h=0}=U\g$. Point 1)
essentially
says that it is possible to find $h$-dependent functions of the
generators
of $U\g$ which satisfy the algebra relations of the Drinfel'd-Jimbo
 generators
of $\uqg$ and vice versa.

We recall here that the quasi-triangular Hopf algebras
$\uqg$ can be obtained as quantizations of Poisson-Lie groups
associated with solutions of the modified classical
Yang-Baxter equations (MCYBE) corresponding to $\g$.

If the Hopf algebra $H$ can be obtained as the quantization of a
Poisson-Lie group associated with a solution
of the classical Yang-Baxter equation (CYBE) corresponding to
some
$\g$,\footnote{In this case $H$ is is triangular,  \ie
$\R_{21}\R_{12}={\bf 1}$}
then another (and chronologically preceding)
theorem by Drinfel'd \cite{drinf} states the existence of a different
$\F$
with similar properties as in the previous theorem---except that
now it is
enough to twist  $(U \g)([[h]])$ equipped with the  ordinary
{\it coassociative} structure in order to obtain $H$. The
quasi-coassociative structure $\Phi$ and the quasi-triangular
structure
$\R_\Phi$ of  point 3) in the theorem reduce to
$\Phi={\bf 1}\ot {\bf 1} \ot {\bf 1}$, $\R_\Phi={\bf 1}\ot {\bf 1}$ ;
the universal $\R$ is given by $\R= \F_{21} \F^{-1}$.
A simple introduction to these topics can be found for instance in
Ref. \cite{Taktadjan}.

As shown in Ref. \cite{jurco}, one can
always choose a unitary $\F$, if $H$ is a compact section of  $U_q(g)$
(\ie when $q\in {\bf R}$). If the $\F$ one starts with is not unitary,
when one simply multiplies it with the invariant (!) tensor $(\F^*
\F)^{-1/2}$
to obtain a new $\tilde{\F}$ that is unitary.

These theorems suggest that one can use the unitary twisting 
operator $\F$ to build
$F_{12}$ for a 2-particle sytem.
For example:
\begin{enumerate}
\item If $\A =\rho( \uqg)$, then we choose
$$ F =\rho^{\ot 2}( \F) .$$
\item If $\A =$ classical Heisenberg algebra
$\otimes U^{\mbox{spin}}(su(2))
\otimes \rho(\uqg)$, were $\uqg$ plays the role of an internal
symmetry,
then we can set
$$ F_{12} = \id^{(2)}_{\mbox{Heisenberg}} \otimes 
\id^{(2)}_{\mbox{spin}}
\otimes \rho^{\ot 2}( \F) $$
\item If $\A$ is the q-deformed Poincar\'e algebra of ref.
\cite{wess2,majid},
and
$H$ is the corresponding q-deformed Lorentz Hopf algebra,
realized through $\rho$ in $\A$, then we can again define
$$ F_{12} = \rho^{\ot 2}( \F),$$
where $\F$ belongs to the homogeneous part.
The same applies for other
inhomogeneous algebras, like the q-Euclidean ones, constructed
from
the braided semi-direct
product \cite{majid} of a quantum space and of the corresponding
homogeneous quantum group. For both of these examples the
one-particle representation theory is known \cite{wess2,fioeu}.
\end{enumerate}

For $n$-particle systems one can set
$F_{12...n}=\rho^{\ot n}( \F_{12...n})$,
where now we should choose one particular element $\F_{12...n}$ of
$H^{\ot n}$ satisfying the condition
\eq
\Delta(x)=\F_{12...n}\Delta_c(x)(\F_{12...n})^{-1}.
\label{condition}
\en
To obtain one such $\F_{12...n}$ it is enough to act on eq.
(\ref{fdf})
 $(n-2)$ times with the coproduct in some arbitrary order.
When $n=3$, for instance, one can use
either $\F'_{123}:=[(\Delta\ot id)(\F)]\F_{12}$ or
$\F''_{123}:=[(id\ot\Delta)(\F)]\F_{23}$. These two elements
coincide in
the case previously mentioned
of Hopf algebras associated to solutions of the CYBE, as proved by
Drinfeld \cite{drinf}. In the the case of $\uqg$, they do not
coincide, but
nevertheless
$\Phi:=\F''_{123}(\F'_{123})^{-1}\neq {\bf 1}\ot{\bf 1}\ot{\bf 1}$
commutes with $\Delta^{(2)}(H)$. In section (\ref{example}) we
will show
(in the $U_q(su(2))$ case) how to find a continuous family of
$\F_{123}$ interpolating between $\F'_{123}$ and $\F''_{123}$.

\paragraph{Note:}
{}From (\ref{fdf}) follows  $(\tau\circ\Delta)(a) =
{\cal M} \Delta(a)
{\cal M}^{-1}$ with ${\cal M} := \F_{21} \F^{-1}.$
This is not the usual relation $(\tau\circ\Delta)(a) = \R \Delta(a)
\R^{-1}$ of
a quasi-triangular Hopf algebra; the latter is rather obtained by
rewriting
equation (\ref{fdf}) in the form
$\Delta(a) = \F q^{t/2} \Delta_c(a) q^{-t/2} \F^{-1}$
where $t = \Delta_c(C_c) - 1\ot C_c  - C_c \ot 1$ is the invariant
tensor \
($[t , \Delta_c(a) ] = 0 \x\forall\x a \in U \g$) \  corresponding to
the Killing metric, and $C_c$ is the quadratic casimir of $U\g$.
 ${\cal M}$, unlike $\R$, has not nice properties
under the coproducts $\Delta\ot id$, $id\ot \Delta$.
The reader might wonder whether we could use  equation
$[P_{12}R, (\A\ot\A)'_+]=0$ (where $R=\rho^{\ot 2}(\R)$),
instead of eq. (\ref{commop}), to single out a
modified symmetric algebra $(\A\ot\A)'_+ \subset \A\ot\A$; in fact,
the former is also an equation fulfilled by
$\rho^{\ot 2}(\Delta(H))$
and reduces to the classical eq. (\ref{symop})  in the limit
$q\rightarrow 1$.
However $[P_{12}R, (\A\ot\A)'_+]=0$ is fulfilled {\it only} by the
sub-algebra
$\rho^{\ot 2}(\Delta(H))\subset(\A\ot\A)$ itself, essentially because
$q^{t / 2}$ does not commute with {\em all}\/ symmetric operators,
but only with the ones corresponding to coproducts. Therefore, 
$(\A\ot\A)_+'$ defined via $P_{12}R$ (instead of $P_{12}^{F_{12}}$) 
is not big enough to be
in one-to-one correspondence with  the classical $(\A\ot\A)_+$, \ie
is not suitable for our purposes.

Explicit universal $\F$'s for  $\uqg$ are not given in the literature,
up to our knowledge; an explicit universal $\F$ for a family of
deformations  (which include quantizations of solutions of
both of a CYBE and of a MCBYE) of the Heisenberg group in one
dimension was given in Ref.  \cite{bonechi}.

However, for most practical purposes one has to deal with
representations $F$ of $\F$.  A general method for constructing
the matrices $F$ acting on tensor products of two arbitrary
irreducible representations of compact sections of $\uqg$ is
presented
in Ref. \cite{eng}---there explicit formulas are given for 
the $A,B,C,D$-series
in the fundamental representation.
In \cite{cuza} matrices twisting the classical coproduct into the $q$-deformed
one were constructed from $q$-Clebsch-Gordan coefficients.

Moreover, in the intrinsic formulation of the twisted
(anti-)symmetrization
postulates [eqs. (\ref{astate}) -- (\ref{asop})] one only needs 
the twisted
permutators $P_{12\ldots n}^{F_{12\ldots n}}$
(not the $F_{12\ldots n}$ themselves); explicit universal
expressions for the
latter can be found much more easily, as we show in section
\ref{example}
for $P_{12}^{\tiny\F_{12}}$ in the case $H=U_q(su(2))$.

We conclude that the quantum symmetry is compatible with identical particle
{\em states}\/ in the twisted multi-particle description.

\subsubsection{Transformation of Operators}
\label{operators}

Now we want to see if a consistent transformation of the
twisted-symmetric operators  can be defined.

As we have
seen in section \ref{onep}, the action on one-particle operators
which makes
eq. (\ref{actsta3}) consistent with eq. (\ref{actsta2}) looks formally
like  the quantum adjoint
action. A subtle but  important change in the
definition of the action on multi-particle operators  is needed
in order to reach the same goal for multi-particle systems.
Our task in this section is twofold: first we have to find
the right
action of the Hopf algebra $H$ for tensor products of $\A$,
then we have to show that the definition of ``twist symmetric''
operators (associated to identical particles) is
stable under this action. As before, we assume
that $\rho$ is a unitary representation that
realizes the Hopf algebra $H$ of transformations in $\A$.

Let $\Op^{(n)} \in \A^{\ot n}$ (or a properly
symmetrized subspace), $|\psi_n\ra\in\Hil^{\ot n}$ (or a properly
(anti)symmetrized subspace); we require, as in the one-particle
case,
\eq
(x_{(1)} \trs \Op^{(n)}) (x_{(2)} \tr |\psi_n\ra ) = x \tr \big(\Op^{(n)}
|\psi_n\ra \big)
=  \rho^{(n)}(x) \Op^{(n)} |\psi_n\ra.
\label{actsta4}
\en
Recalling eq. (\ref{actstam})
it is easy to see that to satisfy this goal
the action (\ref{adjoint}) has to
generalize to multi-particle operators in the following way:
\eq
\ar
x \trs \Op^{(n)} & = & \rho^{(n)}(x_{(1)})\: \Op^{(n)}
         \:\rho^{(n)}(S x_{(2)})\\
   & = & \rho^{\ot n} \big(x_{(1)}\ot \ldots \ot x_{(n)}\big)
   \:\Op^{(n)} \:\rho^{\ot n}\big(S x_{(2n)}\ot
   \ldots \ot S x_{(n+1)}\big).\label{symmetry}
\an
\en
{\em Remark:}\ In the case that $\Op = \rho(y)$ with $y \in H$ the
action
on one-particle operators is nothing but the
adjoint action $x \ad y = x_{(1)} y S(x_{(2)})$. The action on
multi-particle
operators is however different: For instance in the case that $
\Op^{(2)} =
(\rho\ot\rho)(y_i\ot y^i)$ with $y_i\ot y^i \in H \ot H$ we get
$$x \trs (y_i\ot y^i) = x_{(1)} y_i S x_{(4)} \ot x_{(2)} y^i S x_{(3)}$$
and {\em not}\/
$$
x \ad (y_i\ot y^i) = x_{(1)} \ad y_i \ot x_{(2)}\ad y^i
=  x_{(1)} y_i S x_{(2)} \ot x_{(3)} y^i S x_{(4)}
$$
as one might have expected. Both actions ``$\ad$'' and ``$\trs$''
coincide
for co-commutative coproducts.
The former action treats multi-particle operators as tensor products
of $H$-modules, the latter action is related to the natural Hopf
algebra
structure on $\Delta(H)$ that is given in Sweedler's book
\cite{Sweedler}.
Briefly, Sweedler's argument is the following. For any given
number $n$,
$\Delta^{(n-1)}(H)$ can be viewed as a Hopf algebra with a
natural
coproduct. Now formula (\ref{adjoint}) is applicable for any
$n$---we just have to take care to use the natural Hopf algebra
structure
for each of the $\Delta^{(n-1)}(H)$.\footnote{The action ``$\trs$''
was also
used in Ref.~\cite{SWZ} to define covariance properties of
tensors in
$H^{\ot n}$}

The notion of unitary multi-particle transformations generalizes to
$n$ particles  in an obvious way,
\eq
(u \trs \Op^{(n)})^* = u \trs (\Op^{(n)})^* \qquad \forall \Op^{(n)}
\in A
\en
and again is satisfied if $u^* = S(u)$.

We now want to show that the transformation we have found is
compatible
with the symmetrization of operators in the twisted
multi-particle description.
\paragraph{Symmetric coproduct} First consider the co-commutative case. Let
$$
(\A\ot\ldots\ot\A)_+ = \{ a \in \A\ot\ldots\ot\A : [P_{i,i+1} , a] =0 ,
i=1,\ldots n-1\}
$$
be the completely symmetrized space of $n$-particle operators.
In the case of a co-commutative \ie {\em
symmetric}\/ coproduct $\Delta_c$
any of the permutation operators $P_{i,i+1}$ will commute with the
action (\ref{symmetry}):
\eq
\ar
\Big[P_{i,i+1} \, , \, \big(x \trs \Op^{(n)}\big) \Big]
& = & \Big[ P_{i,i+1} \, , \,\rho^{\ot n} \Big(
\Delta_c^{(n-1)}(x_{c(1)}) \Big)
   \:\Op^{(n)}\: \rho^{\ot n}\Big(\Delta_c^{(n-1)}(S_c x_{c(2)})\Big)\Big]\\
& = & x \trs \big[P_{i,i+1} \, , \,\Op^{(n)}\big], \qquad \mbox{for }
\Delta_c \mbox{ cocommutative }.
\an \label{cococ}
\en
Here $x_{c(1)}\ot x_{c(2)}\equiv\Delta_c(x)$ and $S_c$ 
is the cocommutative antipode.

\paragraph{Deformed coproduct} Let ${\F_{12\ldots n}}\in H^{\ot n}$ be
as in equation (\ref{condition}) namely 
such that
$\Delta^{(n-1)}(x) = {\F_{12\ldots n}}\Delta_c^{(n-1)}(x)
{\F_{12\ldots n}}^{-1}$ for all $x \in H$.
As in the previous section
we will use its representation  $F_{12\ldots n} \equiv
\rho^{\ot n}({F_{12\ldots n}})$ for the similarity transformation of
section~\ref{twistmult}. 
We note that relation (\ref{cococ}) also holds with $x_{c(1)}\ot x_{c(2)}$
and $S_c$ replaced by the non-cocomutative $x_{(1)}\ot x_{(2)}\equiv\Delta(x)$
and the corresponding antipode $S$:
\eq
\ar
& & \Big[ P_{i,i+1} \, , \,\rho^{\ot n} \big(
\Delta_c^{(n-1)}(x_{(1)}) \big)
   \:\Op^{(n)}\: \rho^{\ot n}\big(\Delta_c^{(n-1)}(S x_{(2)})\big)\Big]\\
& & = \rho^{\ot n} \big(\Delta_c^{(n-1)}(x_{(1)}) \big) 
\big[P_{i,i+1} \, , \,\Op^{(n)}\big]
\rho^{\ot n}\big(\Delta_c^{(n-1)}(S x_{(2)})\big). 
\an \label{mcococ}
\en
Conjugating this relation by $F_{12\ldots n}$
we easily find the non-cocommutative analog of equation 
(\ref{cococ}), because $\rho^{(n)}(x) = 
\rho^{\otimes n}(\Delta^{(n-1)}(x)) = F_{12\ldots n} 
\rho^{\otimes n}(\Delta_c^{(n-1)}(x))
F^{-1}_{12\ldots n}$:
\eq
\Big[P^{F_{12\ldots n}}_{i,i+1} \, , \, \big(x \trs \Op^{(n)}\big) \Big]
= x \trs \big[P^{F_{12\ldots n}}_{i,i+1} \, , \,\Op^{(n)}\big] \qquad
\forall x \in H.
\label{noncococ}
\en
Consequently, since the LHS vanishes if the RHS does:
\eq
H\,:\: (\A\ot\ldots\ot\A)_+^{F_{12\ldots n}}\:\rightarrow\:
(\A\ot\ldots\ot\A)_+^{F_{12\ldots n}}.
\en

The quantum symmetry is hence compatible with identical
particle {\em operators}\/
in the twisted multi-particle description.\\[1em]
{\em Remark:}\ The transformation (\ref{symmetry}) is not the
only one \label{rem}
compatible with the twisted symmetrization. The important point is that the
transformation must be based on $\rho^{(n)}(x) = \rho^{\otimes
n}(\Delta^{(n-1)}(x))$. The ordinary
commutator
$\big[\rho^{(n)}(x)\, , \, \Op^{(n)}\big]$ also leaves
$(\A^{\ot n})^{F_{12\ldots n}}_+$ invariant, simply because
$\rho^{(n)}(x) \in
(\A^{\ot n})^{F_{12\ldots n}}_+$. These two transformations
usually
coincide in ordinary quantum mechanics. Here they have different
interpretations:
Let $h \subset H$ be a sub-algebra of $H$.
The operator $\Op^{(n)}$, $n\ge 1$,
is symmetric (\ie invariant) under the transformations
generated by $x\in h$ if
\eq
x \trs \Op^{(n)} = \Op^{(n)} \epsilon(x);
\label{invar}
\en
it may be simultaneously diagonalizable with elements in $h$ if
\eq
\big[\rho^{(n)}(x)\, , \, \Op^{(n)}\big] = 0.
\label{commu}
\en
The two properties coincide if $\Delta(h) \subset h \ot H$.
This can be seen as follows:
\eqa
\rho^{(n)}(x) \Op^{(n)}|\psi_n\ra & \stackrel{(\ref{actstam})}{=}&
x\trs (\Op^{(n)}|\psi_n\ra)   \nonumber\\
\stackrel{(\ref{actsta4})}{=}
(x_{(1)}\trs \Op^{(n)})(x_{(2)}\trs|\psi_n\ra) &
\stackrel{(\ref{invar})}{=}
&\varepsilon(x_{(1)})  \Op^{(n)}(x_{(2)}\trs|\psi_n\ra) \\
=  \Op^{(n)} (x\trs|\psi_n\ra) & = & \Op^{(n)} \rho^{(n)}(x)|\psi_n\ra
\nonumber
\ena
for any $|\psi_n\ra\in \Hil^{\ot n}$, so that eq. (\ref{invar}) implies
eq (\ref{commu}); in the same way one proves the converse.
The physical relevance of this case is self-evident: if both
$\Op^{(n)} $ and $\rho^{(\ot n)}(x)$ are hermitean,
then they can be diagonalized simultaneously; if one of the two,
say $\rho^{(\ot n)}(x)$, is not hermitean, given an eigenvector
$|\psi_n\ra$ of $\Op^{(n)}$, $\rho^{(\ot n)}(x)|\psi_n\ra$ will be
another
belonging to the same eigenvalue.

\sect{Explicit Example: $H=U_q(su(2))$}
\label{example}

We consider as a simple example of a one-particle quantum 
mechanical system transforming under a quantum group action
the case of a q-deformed rotator, 
$\A\equiv \rho(H):=\rho[U_q(su(2))]$,
with $q\in {\bf R}^+$. We determine the twisted symmetry of 
the systems consisting of $n\ge 2$ particles of the same kind.

\subsection{$n=2$ particles}
\label{example1}

Let us first assume that 
the states of the system belong to an irreducible 
$*$-repre\-sent\-ation 
of $H$, namely 
$\Hil\equiv V_j$, where $V_j$ denotes the highest weight 
representation of  
$U_q(su(2))$ with highest weight $j=0,\frac 12,1, ...$. It is very 
instructive to find 
out what $(\Hil\ot \Hil)_{\pm}^{F_{12}}$ and 
$(\A\ot \A)_+^{F_{12}}$ in this 
example are.

According to point 1. of  the Drinfel'd theorem, we can 
identify $U_q(su(2))$ 
and $U(su(2))$ as algebras; therefore, $V_j$ can be thought as 
the representation 
space of either one. Similarly, $V_j\ot V_j$ can be considered as 
the carrier space 
of a (reducible) representation space of either  
$U_q(su(2))\ot U_q(su(2))$ 
or $U(su(2))\ot U(su(2))$; moreover, 
$F_{12}(V_j\ot V_j)=V_j\ot V_j$. Thus,
we can decompose it  into irreducible components either of 
$U_q(su(2))$ or $U(su(2))$, the 
operators on it being defined as 
$\rho^{(2)}(X)=\rho^{\ot 2}[\Delta(X)]$ or 
$\rho^{(2)}_c(X)=\rho^{\ot 2}[\Delta_c(X)]$ respectively:
\eq
V_j\ot V_j=\cases{\bigoplus\limits_{0\le l\le j}\V_{2(j-l)}^q\oplus
\bigoplus\limits_{0\le l\le j-\frac 12}\V_{2(j-l)-1}^q   \cr
\bigoplus\limits_{0\le l\le j}\V_{2(j-l)}\oplus
\bigoplus\limits_{0\le l\le j-\frac 12}\V_{2(j-l)-1};   \cr}
\en
here $\V^q_J$ (resp. $\V_J$) denotes the irreducible component 
of 
$U_q(su(2))$ (resp. $U(su(2))$) with highest weight $J$. 
Moreover, 
from point 2) of the theorem it follows
\eq
F_{12}\V_J=\V_J^q, 
\label{vtwist}
\en

Let us recall now that the $\V_J$'s have well-defined symmetry 
w.r.t the permutation, namely
$\V_{2j}, \V_{2(j-1)}, \ldots$ are symmetric, 
$\V_{2j-1},\V_{2j-3},...$ are 
antisymmetric.
This follows from the fact that $\rho^{(2)}_c(X)$ and $P_{12}$ 
commute. 
Hence
\eqa
& & (V_j\ot V_j)_+=\bigoplus\limits_{0\le l \le j}\V_{2(j-l)}\\  
\nonumber
& & (V_j\ot V_j)_-=
\bigoplus\limits_{0\le l\le j-\frac 12}\V_{2(j-l)-1}. 
\label{pmdecom}
\ena
{}From eq.'s (\ref{vtwist}), (\ref{pmdecom}) we finally find
\eqa
& & (V_j\ot V_j)^{F_{12}}_+:= F_{12}(V_j\ot V_j)_+=
\bigoplus\limits_{0\le l \le 
j}\V_{2(j-l)}^q\\ \nonumber
& & (V_j\ot V_j)^{F_{12}}_-:= F_{12}(V_j\ot V_j)_-=
\bigoplus\limits_{0\le l\le 
j-\frac 12}\V_{2(j-l)-1}^q.  
\label{pmqdecom}
\ena
This equation says that the subspaces 
$\V^q_J\subset V_j\ot V_j$ have 
well-defined ``twisted symmetry''. We can use it to 
build $(V_j\ot V_j)^{F_{12}}_{\pm}$ recalling how the 
representations 
$\V^q_J$ are obtained. 
For this scope, we just have to recall the explicit algebra 
relations and coproduct of the generators $h,X^{\pm}$
of $U_q(su(2))$:
\eqa
[h,X^{\pm}]& = & \pm 2 X^{\pm} \qquad \qquad \qquad \qquad 
  [X^+,X^-] =  \frac{q^{h}-q^{-h}}{q- q^{-1}}  \nonumber\\
\Delta(h) & = & {\bf 1}\ot h + h\ot {\bf 1} \qquad \qquad 
\Delta(X^{\pm})  =  X^{\pm}\otimes q^{-\frac{h}2} + q^{\frac{h}2}
\otimes X^{\pm}.
\label{uqsu2comrel}
\ena
Let $\{|j,m\ra\}_{m=-j,1-j,...j}$ be an orthonormal basis of $V_j$ 
consisting of 
eigenvectors of $\rho(\frac h2)$ with eigenvalues $m$.  The
generators $X^{\pm}$ can be represented in terms of this basis 
in the following way
\eqa
\rho(X^+)|j,m\ra&=&\sqrt{[j+m+1]_q[j-m]_q}|j,m+1\ra,\\
\rho(X^-)|j,m\ra&=&\sqrt{[j-m+1]_q[j+m]_q}|j,m-1\ra,
\ena
where $[x]_q:=\frac{q^x-q^{-x}}{q-q^{-1}}$.
As well known, the 
highest weight vector $\Vert J,J\ra\in \V^q_J$---from  
which the whole representation $\V^q_J$ can be generated 
by repeated 
applications of $\rho^{(2)}(X^-)$---is
obtained by solving the equation 
$\rho^{(2)}(X^+)\Vert J,J\ra=0$ for the 
coefficients $a_h$ of the general ansatz
\eq
\Vert J,J\ra=\sum\limits_{h=\mbox{\scriptsize 
max}\{-j,J-j\}}^{\mbox{\scriptsize min}\{j,J+j\}} 
a_h |j,h\ra\ot |j,J-h\ra.
\en

Now we are ready to understand the difference between 
$(H\ot H)_+^{F_{12}}$ and 
its sub-algebra
$\rho^{(2)}(H)$: 
\eqa
\rho^{(2)}(H)&\ni& a:\V^q_J\rightarrow \V^q_J,\\
(H\ot H)_+^{F_{12}}&\ni& 
b:(V_j\ot V_j)^{F_{12}}_{\pm}
\rightarrow (V_j\ot V_j)^{F_{12}}_{\pm}.
\ena
The elements of 
$[\rho(H)\ot \rho(H)]_+  \setminus \rho^{(2)}(H)$ will in general 
map $\V^q_J$ out of itself, into some $\V^q_{J'}$ with $J'\neq J$.

If $\Hil$ carries a reducible  $*$-representation of $H$, it will be 
possible to 
decompose it into irreducible representations  $V_j$,
\eq
\Hil=\bigoplus\limits_{j\in {\cal J}}V_j \qquad\qquad {\cal J}\subset 
\nn_0/2:=\{0,\frac 12,1\ldots\}; 
\en
then 
\eq
\Hil\ot \Hil=\bigoplus\limits_{j_1,j_2\in {\cal J}}V_{j_1}\ot V_{j _2},
\label{decom}
\en
and each $V_{j_1}\ot V_{j _2}$ itself will be a representation. 
If $j_1=j_2$, the 
considerations above apply. If $j_1\neq j_2$, the irreducible 
components 
$\V^q_J$  ($J=|j_1-j_2|,|j_1-j_2|+1,\dots, j_1+j_2$) contained in 
$V_{j_1}\ot V_{j _2}$ 
of course  {\it will not} have well-defined symmetry (neither 
classical nor 
twisted) under permutations. However, the 
irreducible components  $\tilde{\V}^q_J$ contained in 
$V_{j_2}\ot V_{j _1}$ 
will be characterized by the same set of highest weights $J$.
One can split
$\V^q_J\oplus \tilde{\V}^q_J$, and therefore 
$V_{j_1}\ot V_{j_2}\oplus V_{j_2}\ot V_{j_1}$,
into the direct sum of one (twisted) symmetric and one (twisted)
antisymmetric components
\eq
\left[V_{j_1}\ot V_{j_2}\oplus V_{j_2}\ot V_{j_1}\right]^
{F_{12}}_{\pm}=
F_{12}\frac 12[{\bf 1} \pm P_{12}]\left[V_{j_1}\ot V_{j_2}\oplus 
V_{j_2}\ot V_{j_1}\right]_{\pm}
\en
(the symbol $F_{12}$ has to be dropped in the untwisted case).
Let $\{\Vert J,M\ra^q_{12}\}_{M=-J,...,J}$ be an orthonormal 
basis 
of $\V^q_J$ consisting of eigenvectors of 
$\rho^{(2)}(h)$ and of $\rho^{(2)}(C_q)$ ($C_q$ denotes the 
casimir),
and let 
\eq
\Vert J,M\ra^q_{12}:=\sum\limits_{m_1,m_2}{\cal C}^{j_1,j_2}_
{m_1,m_2}
(J,M,q)|j_1,m_1\ra|j_2,m_2\ra
\en
be the explicit decomposition of $\Vert J,M\ra^q_{12}$ in the 
tensor
product basis of $V_{j_1}\ot V_{j _2}$. Then the set
$\{\Vert J,M\ra^q_{21}\}_{M=-J,...,J}$ with
\eq
\Vert J,M\ra^q_{21}:=\sum\limits_{m_1,m_2}{\cal C}^{j_2,j_1}_
{m_2,m_1}
(J,M,q)|j_2,m_2\ra|j_1,m_1\ra,
\label{basis}
\en
will be an orthonormal basis of $\tilde{\V}^q_J$ consisting of 
eigenvectors of 
$\rho^{(2)}(h)$ and of the casimir $\rho^{(2)}(C_q)$ with the same 
eigenvalues.
Defining
\eq
\Vert J,M\ra^q_{\pm}:=N\left( \Vert J,M\ra^q_{12}\pm 
\Vert J,M\ra^q_{21}\right),
\qquad\qquad\qquad N^{-1}:=\sqrt{2}
 \en
we can easily realize that $\{\Vert J,M\ra^q_{\pm}\}_{J,M}$  is an  
orthonormal basis of 
$(V_{j_1}\ot V_{j_2}\oplus V_{j_2}\ot V_{j_1})^{F_{12}}_{\pm}$.

Note that, if $j_1=j_2\equiv j$ and we set $N^{-1}=2$ in formula 
(\ref{basis}),
then the vectors $\Vert J,M\ra^q_+$ will make up 
the same
orthonormal basis of  $V_j\ot V_j$
as before
(they will have twisted symmetry $(-1)^{J-2j}$, see the previous 
case)
whereas the vectors $\Vert J,M\ra^q_-$ will vanish.

We are now ready to find, as announced in sections \ref{twistmult},
\ref{qsym}, the ``universal  twisted
permutator'' $P_{12}^\ef$ of $U_q(su(2))$, defined 
through
the property that the twisted permutation operator 
$P_{12}^{F_{12}}$ 
on any tensor product $V\ot V$ [$V$ being the carrier space of
a representation $\rho$ whatever of $U_q(su(2))$] can be obtained
by $P_{12}^{F_{12}}=\rho^{\ot 2} (P_{12}^\ef)$.

We decompose $V\ot V$ as in formula (\ref{decom}). The casimir of
$U_q(su(2))$
\eq
C_q=X^-X^++ \left(\frac{q^{\frac{h+1}2}-q^{\frac{-h-1}2}}
{q-q^{-1}}\right)^{2}
\en
has eigenvalues $([j+\frac 12]_q)^2$; in the limit 
$q\rightarrow 1$: 
$C_q\rightarrow C_c+\frac14$, where $C_c$ is 
the usual casimir
of $U(su(2))$ with eigenvalues $j(j+1)$. Defining $f(z)$ by
\eq
\log_q[f(z)]:=\left\{\frac 1{\ln(q)}\sinh^{-1}\left[\frac{(q-q^{-1})
\sqrt{z}}2\right]\right\}^2-\frac 14,
\en
it is easy to verify that $f(C_q)$ has eigenvalues $q^{j(j+1)}$.
Let $\hat R:=P_{12}[\rho^{\ot 2} (\R)]$.  Recalling the formula 
$\R={\cal F}_{21}q^{\frac t2}{\cal F}_{12}^{-1}$, we realize that 
the vectors
$\Vert J,M\ra^q_{\pm}\in (V_{j_1}\ot V_{j_2}\oplus 
V_{j_2}\ot V_{j_1})^{F_{12}}_{\pm}$ ($j_1\neq j_2$) are 
eigenvectors of
$\rho^{\ot 2}\left[f({\bf 1}\ot C_q)f(C_q\ot {\bf 1})\left[
f(\Delta(C_q))\right]^{-1}\right]\hat R$ and 
$P_{12}^{F_{12}}$ with the same eigenvalue $\pm 1$. If
$j_1=j_2=j$, the same holds for the vectors $\Vert J,M\ra^q_+$ 
(which form a basis of $V_j\ot V_j$). Since this holds for all
$j_1,j_2$ appearing in the decomposition (\ref{decom}), and 
if we let $j_1,j_2$ range on ${\cal J}$ the above vectors
make up a basis of $V\ot V$, then 
\eq
P_{12}^{F_{12}}=f({\bf 1}\ot\rho(C_q))f(\rho(C_q)\ot {\bf 1})
\left[f\left(\rho^{(2)}(C_q)\right)\right]^{-1}\hat R
\en
on $V\ot V$. We prefer to rewrite $\hat R$ as 
$\hat R=[\rho^{\ot 2} (\R_{21})]P_{12}$, where $\R_{21}=\tau(\R)$
and $\tau$ is the abstract permutator. 
Since this equation holds for an arbitrary representation $\rho$, 
we 
can drop the latter symbol and obtain the 

\paragraph{Universal expression for the twisted
permutation operator of $U_q(su(2))$:}
\eq
P_{12}^\ef=f({\bf 1}\ot C_q)f(C_q\ot {\bf 1})\left[f(
\Delta(C_q))\right]^{-1}\R_{21}\circ \tau
\en
We omit here the well-known expression for the universal $\R$  
\cite{dr}.

\subsection{$n\ge 3$ particles}

When $n\ge 3$, for any given space $V$ the decomposition of 
$\bigotimes^n V$ into irreducible representations  of
the permutation group contains components with partial/mixed 
symmetry,
beside the completely symmetric and the completely 
antisymmetric ones.
\footnote{The Young tableaus  provide the rules for finding the 
complete 
decomposition for any $n$.}
If $n=3$, for instance, some components can be diagonalized 
{\it either}  
w.r.t. to $P_{12}$ {\it or} w.r.t. $P_{23}$ (but not w.r.t. both of them
simultaneously).
If $n=4$, all components can be diagonalized simultaneously w.r.t. 
$P_{12}$ and $P_{34}$, and some will have mixed symmetry (e.g. 
will be
symmetric in the first pair and antisymmetric in the second, or 
vice versa). 
We recall that the explicit knowledge of components with 
mixed/partial 
symmetry  is required to build $(\Hil^{\ot n})_{\pm}$ 
if the Hilbert space $\Hil$ of one particle is the tensor product of 
different
spaces, $\Hil=V\ot V'$, as in example~2 in subsection~\ref{transta}.

It is easy to
realize that similar statements hold in the case of the twisted 
symmetry.

Let us consider again the case $V_j$, and let $n=3$ for the sake of 
simplicity. We show how to construct two different orthonormal 
bases of 
$V_j\ot V_j\ot V_j$ with (partial) symmetry, and a continuous 
family 
of $F_{123}$ on $V_j\ot V_j\ot V_j$.

There is evidently only one irreducible representation
with highest weight $J=3j$, the highest weight vector being
$|j,j\ra|j,j\ra|j,j\ra$. But there are two independent irreducible 
representations
with highest weight $J=3j-1$, e.g. those having highest weight 
vectors 
$\frac1{\sqrt{2}}\left(|j,j-1\ra|j,j\ra\pm|j,j\ra |j,j-1\ra\right)|j,j\ra$.
The latter are symmetric and antisymmetric respectively w.r.t. 
$P_{12}$,
but are mixed into each other by the action of $P_{23}$; 
alternatively, 
one can combine these two representations into two new ones,
having highest weight vectors 
$\frac1{\sqrt{2}}|j,j\ra\left(|j,j-1\ra|j,j\ra\pm|j,j\ra |j,j-1\ra\right)$,
which are symmetric and antisymmetric respectively w.r.t. 
$P_{23}$,
but are mixed into each other by the action of $P_{12}$. One can 
easily 
verify that the first two representations are eigenspaces of
$\rho^{(2)}_c(C_c)\ot id$ with eigenvalues $(2j\pm \frac 12)^2$, 
the latter
two are  eigenspaces of $id \ot \rho^{(2)}_c(C_c)$ with the same 
eigenvalues.
The operators $\rho^{(3)}_c(C_c),\rho^{(3)}_c(h)$ and either 
$\rho^{(2)}_c(C_c)\ot id$ or $id \ot \rho^{(2)}_c(C_c)$ make up a 
complete set of commuting observables over $V_j\ot V_j\ot V_j$. 
Let
\eq
\{\Vert J,M,r\ra_{12}\}_{J,M,r}, \qquad [\mbox{respectively:}\quad  
\{\Vert J,M,s\ra_{23}\}_{J,M,s}],
\en
with
\eq
 j\le J\le 3j, \z  -J\le M\le J,\z
\mbox{max}\{0,j\!-\!J\}\le r,s\le \mbox{min}\{2j,j\!+\!J\}
\label{prima}
\en
denote an orthonormal basis of eigenvectors of 
$\rho^{(3)}_c(C_c)$, $\rho^{(3)}(h)$ and
$\rho^{(2)}_c(C_c)\ot id$ [respectively:\x $id \ot \rho^{(2)}_c(C_c)$] with 
eigenvalues
$J(J+1),M$ and $r(r+1)$ [respectively:\x $s(s+1)$].
In particular, 
\eqa
\Vert 3j-1,3j-1,2j-\frac 12 \pm\frac 12 \ra_{12} = 
\frac1{\sqrt{2}}\left(|j,j-1\ra|j,j\ra\pm|j,j\ra |j,j-1\ra\right)
|j,j\ra\nonumber\\
\Vert 3j-1,3j-1,2j-\frac 12 \pm\frac 12 \ra_{23} = 
\frac1{\sqrt{2}}|j,j\ra\left(|j,j-1\ra|j,j\ra\pm|j,j\ra |j,j-1\ra\right)
\ena
It is easy to verify that in general the subspace of 
$V_j\ot V_j\ot V_j$ which is anti-symmetric/symmetric w.r.t.\ $P_{12}$
is spanned by the vectors $\Vert J,M,r\ra_{12}$
with $r-\mbox{min}\{2j,j\!+\!J\}$ odd/even, 
and similarly for $P_{23}$.

For fixed $J,M$, there exists a unitary matrix $U(J)$ such that
\eq 
\Vert J,M,s\ra_{23}=U(J)_{sr}\Vert J,M,r\ra_{12}
\label{ultima}
\en

Formulae formally identical to eqs. (\ref{prima}), (\ref{ultima}) 
hold when 
$q\neq 1$;  we will introduce an additional index $q$ in all objects 
to denote this dependence.

The elements of $\rho^{(3)}(U_q(su(2)))$ in these two bases read
\eq
\rho^{(3)}(X)=\cases{\sum\limits_J\sum\limits_r\sum\limits_{M,M'}
X_{M,M'}(J) \Vert J,M,r,q\ra_{12}~ _{12}\la J,M,r,q\Vert\cr
\sum\limits_J\sum\limits_s\sum\limits_{M,M'}
X_{M,M'}(J) \Vert J,M,s,q\ra_{23} ~_{23}\la J,M,s,q\Vert, \cr}
\en
and the matrix elements $X_{M,M'}(J)$ do not depend on $r,s$.

Now it is easy to check that we can find  
many-parameter continuous families of matrices $F_{123}$
satisfying eq.  (\ref{condition}), in the form
\eq 
F_{123}=\cases{\sum\limits_J\sum\limits_M\sum\limits_r
A_{r,r'}(J) \Vert J,M,r,q\ra_{12}~_{12}\la J,M,r',1\ra\Vert  \cr
\sum\limits_J\sum\limits_M\sum\limits_s
B_{s,s'}(J) \Vert J,M,s,q\ra_{23} ~_{23}\la J,M,s',1\Vert, \cr}
\label{family}
\en
where $A(J)$'s are arbitrary unitary matrices and
$B(J)=[U(J,q)]^*A(J) [U(J,q=1)]^T$. The key
point is that the matrix elements
$A_{r,r'}$ do not depend on $M$, whereas the matrix elements
$X_{M,M'}$ do not depend on $r$. 

It is easy to realize that
the family (\ref{family}) interpolates between the two $F$ matrix 
given
in subsection \ref{transta}, $F_{123}'$ (if we set 
$A_{r,r'}=\delta_{r,r'}$) 
and $F_{123}''$   (if we set $B_{r,r'}=\delta_{r,r'}$).

Considerations analogous to those of subsection \ref{example1} 
can be
done for $n\ge 3$ when $V$ is a reducible representation of 
$U_q(su(2))$.

\section*{Acknowledgments}

We would like to thank R.\ Engeldinger 
for sharing his interpretation of Drinfel'd
twists which helped us in a change of viewpoint on the problem and
we are grateful to S.\ L.\ Woronowicz and W.\ Pusz for pointing 
out Ref. \cite{wopu}.
One of us (G.F.) thanks A.~v.~Humboldt foundation for financial support.
Last but not least we thank J.~Wess for discussion and 
warm hospitality at his institute.

\pagebreak
\sect{Appendix}
\label{appendix}

The following is a short summary of Hopf algebra notions 
{\em relevant to the present article.}
For more detailed discussions of these topics the reader could {\em e.g.}\/
consult  \cite{Sweedler,dr,Drinfeld}.

Hopf algebras can be seen as an abstraction from group algebras and (universal
enveloping algebras of) Lie groups. Taking part of the representation theory
into their very definition, Hopf algebras achieve the unification of such
diverse concepts. 

Mathematically, a Hopf algebra $H$ is an algebra $H(\cdot, + ,k)$ 
over a field $k$
(typically the field of complex numbers) with additional operations $\Delta,
\epsilon, S$ called the coproduct, counit and antipode respectively, satisfying 
suitable axioms.\\
The {\bf coproduct $\Delta$} is an algebra homomorphism,
$\Delta: \: H \rightarrow H \ot H$, that 
fixes the way representations are combined: Let
$\rho, \rho'$ be representations of $H$ on some vector space. The tensor
product $\rho\times\rho'$ of these representations is
\eq
\rho\times\rho' = (\rho \ot \rho') \circ \Delta. \label{tensprodrep}
\en
Let $l \in {\bf g} \subset U{\bf g}$ be a Lie algebra element; its coproduct is
nothing but the well-known angular-momentum addition rule
\eq
\Delta(l) = l \ot 1 + 1 \ot l.\label{exang}
\en
(In the case of quantum groups coproducts are in general not symmetric ({\em 
i.e.} not
co-commutative), see (\ref{uqsu2comrel}).)
To take tensor products of more than two representations we have to apply the
coproduct repeatedly. It does not matter which tensor factors get split again
hereby, since the coproduct is co-associative
\eq
(\Delta \ot i\!d)\Delta(x) = (i\!d \ot\Delta)\Delta(x),\qquad \forall x \in H.
\en
This property is nicely expressed in Sweedlers notation for the coproduct:
($\forall x \in H$)
\eqa
&\Delta(x) \equiv x_{(1)} \ot x_{(2)} \in H \ot H&\\
&(\Delta \ot i\!d)\Delta(x) \equiv x_{(1)}\ot x_{(2)} \ot x_{(3)}
\equiv  (i\!d \ot\Delta)\Delta(x) \in H \ot H \ot H&\\
& \vdots&\nonumber
\ena
Note that (formal) sums of terms are implied in this notation.
Expression (\ref{tensprodrep}) can now be written as:
\eq
(\rho \times \rho')(x) = \rho(x_{(1)}) \ot \rho'(x_{(2)}).
\en
The {\bf counit $\epsilon$}  gives the trivial (1-dimensional) representation:
\eq
\epsilon: \: H \rightarrow k, \qquad \epsilon(x y ) = \epsilon(x) \epsilon(y).
\en
It holds that
\eq
\cdot(\epsilon \ot i\!d) \Delta(x) = x = \cdot( i\!d \ot \epsilon )\Delta(x),
\qquad \forall x \in H,
\en
{\em i.e.}\/ $\epsilon \times \rho = \rho = \rho \times \epsilon$ for all
representations $\rho$. We have $\epsilon(1) =1$, because $\Delta(1) = 1 \ot 1$, 
and
$\epsilon(l) = 0$ for $l \in {\bf g}$ in example (\ref{exang}).\\
The {\bf antipode $S$} is an anti-algebra map 
\eq
S: \: H \rightarrow H, \qquad S(x y ) = S(y) S(x),
\en that defines the contragredient representation $\rho^\vee$ to any given 
representation $\rho$:
\ $\rho^\vee = \rho^T \circ S$ \ (${}^T$ is the transpose).
You may think of $S$ as a generalized inverse.
It holds that 
\eq
S(x_{(1)}) x_{(2)} = 1 \cdot \epsilon(x) = x_{(1)} S(x_{(2)}),
\qquad \forall x \in H,
\en
{\em i.e.}\/ $(\rho^\vee)^T \times \rho = \epsilon = \rho \times
(\rho^\vee)^T$,
and 
\eq
\Delta(S(x)) = S(x_{(2)}) \ot S(x_{(1)}), \qquad \epsilon(S(x)) = \epsilon(x),
\qquad \forall x \in H.
\en
We have $S(1) = 1$ and $S(l) = -l$ for  $l \in {\bf g}$ 
in example (\ref{exang}).\\
Quasi-triangular Hopf-algebras are Hopf-algebras where the non-cocommutativity
is under control by a ``universal'' $\R \in H \ot H$ such that
\eq
x_{(2)} \ot x_{(1)} = \R (x_{(1)} \ot x_{(2)}) \R^{-1},\qquad \forall x \in H,
\label{qybe1}
\en and
\eq
\Delta_1 \R = \R_{13} \R_{23}, \qquad \Delta_2 \R = \R_{13} \R_{12}.
\label{qybe2}
\en
(The subscripts indicate tensor factors here---{\em i.e.}\/ $\Delta_1 \equiv
\Delta \ot i\!d$ and  $\R_{13} \R_{23} =
\sum_{i,j} a_i \ot a_j \ot b_i b_j \in H \ot H \ot H$,
where $\R \equiv \sum_i a_i \ot b_i$, {\em etc.})
It is left as an exercise to the reader to show that (\ref{qybe1})
and (\ref{qybe2}) imply the so-called quantum Yang-Baxter-Equation
\eq
\R_{12} \R_{13} \R_{23} = \R_{23} \R_{13} \R_{12}.
\en
In section~4 we used the concept of quasi-Hopf algebras.
These are in general non-coassociative Hopf algebras with an element
$\Phi \in H \ot H \ot H$ such that
\eq
(i\!d \ot \Delta)\Delta(x) = \Phi (\Delta \ot i\!d)\Delta(x) \Phi^{-1},
\quad \forall x \in H.
\en


\begin{thebibliography}{99}


\bibitem{dr}
V.\ G.\ Drinfeld, {\it Quantum Groups}, Proceedings of the
International
Congress of Mathematicians, Berkeley 1986, Vol. 1, 798.

\bibitem{ji} M.\ Jimbo, Lett.\ Math.\ Phys. {\bf 10} (1986), 63.

\bibitem{frt}
L.\ D.\ Faddeev, N.\ Y.\ Reshetikhin and L.\ A.\ Takhtajan,
{\it Quantization of Lie
Groups and Lie Algebras}, Algebra i Analysis, {\bf 1} (1989),
178; translation: Leningrad Math.\ J.\ {\bf 1} (1990), 193.

\bibitem{wess}
O.\ Ogievetsky, W.\ B.\ Schmidke, J.\ Wess and B.\ Zumino,
Commun.\ Math.\ Phys.\ {\bf 150} (1992), 495.

\bibitem{wess1} J. Wess, B. Zumino, {\em Differential Calculus on
Quantum Planes and Applications}, Talk given on the occasion of the Third
Centenary Celebrations of the Mathematische Gesellschaft Hamburg, March 1990,
KA-THEP-1990-22.

\bibitem{wess2}
M.\ Pillin, W.\ B.\ Schmidke and J.\ Wess, Nucl.\ Phys.\
{\bf B403} (1993), 223.

\bibitem{fio}
G.\ Fiore, Int.\ J.\ Mod.\ Phys.\ {\bf A 8} (1993), 4679; {\em
$SO_q(N)$-isotropic Harmonic Oscillator on the Quantum
Euclidean
space ${\bf R}_q^N$}, LMU-TPW 94-26 and q-alg/9412011.

\bibitem{fioeu}
G.\ Fiore, {\em The Euclidean Hopf algebra $U_q(e^N)$ and its
fundamental Hilbert space representations}, J.\ Math.\ Phys.\ in
press;
{\em The q-Euclidean algebra $U_q(e^N)$ and the corresponding
q-Euclidean lattice}, LMU-TPW 95-4 and q-alg/9505028.

\bibitem{wei} W.\ Weich,
{\em The Hilbert Space Representations for $SO_q(3)$
Symmetric Quantum
Mechanics}, LMU-TPW-1994-5 and hep-th/9404029.

\bibitem{mascho} G. Mack, V. Schomerus, Commun. Math. Phys. {\bf 149} (1992),
513.

\bibitem{drinf}
V.\ G.\ Drinfeld,  Doklady AN SSSR {\bf 273} (1983), 531.

\bibitem{Taktadjan}
L.\ A.\ Takhtajan, {\it Lectures on Quantum Groups}, in
``Introduction to Quantum Groups and Integrable Massive
Models of Quantum Field Theory'', Nankai Lectures on
Mathematical Physics, (World Scientific, 1990).

\bibitem{resh} N.\ Y.\ Reshetikhin, Lett.\ Math.\ Phys.\ {\bf 20} (1990), 331.

\bibitem{luk} J.\ Lukierski, H.\ Ruegg, V.\ N.\ Tolstoy  and  A.\ Nowicki,
{\it Twisted Classical Poincar\'e Algebras},
J.\ Phys. {\bf A 27} (1994), 2389. 

\bibitem{podles}
P.\ Podles,  S.\ L.\ Woronowicz,
{\em On the Classification of Quantum Poincar\'e
Groups},  HEPTH-9412059 and hep-th/9412059;
{\em On the structure of Inhomogeneous Quantum Groups},
HEPTH-9412058 and hep-th/9412058.

\bibitem{Drinfeld}
V.\ G.\ Drinfeld, {\it Quasi Hopf Algebras}, Leningrad Math.\ J.\
{\bf 1} (1990),
1419.

\bibitem{voza} S.\ Vokos, C.\ Zachos, {\it Thermodynamic q-distributions that 
aren't},
Mod.\ Phys.\ Lett. {\bf A 9} (1994), 1.

\bibitem{nelson} C. A. Nelson, M. H. Fields, Phys. Rev. {\bf A 51} (1995), 2410.

\bibitem{zachos} C. Zachos, Mod. Phys. Lett. {\bf A7} (1992), 1559.

\bibitem{zhang} R. Zhang, Lett. Math. Phys. {\bf 25} (1992), 317.

\bibitem{wopu} W. \ Pusz, S. \ L. \ Woronowicz, {\it Twisted
Second Quantization},
Reports on Mathematical Physics {\bf 27} (1989), 231.

\bibitem{eng}
R.\ Engeldinger, {\it On the Drinfel'd-Kohno Equivalence of Groups
and Quantum Groups}, Preprint LMU-TPW 95-13 and q-alg/9509001.

\bibitem{engnote} R.\ Engeldinger, unpublished note (1994) and private
communication.

\bibitem{thesis}
P.\ Schupp, Ph.D.\ thesis, UC Berkeley (1993);
UMI-94-30673-mc (micro fiche), LBL-34942 and hep-th/9312075.

\bibitem{jurco} B.~Jurco, Commun.\ Math.\ Phys.\ {\bf 166} (1994), 63.

\bibitem{majid}
S.\ Majid, J.\ Math.\ Phys. {\bf 34} (1993), 2045.

\bibitem{bonechi}
F.\ Bonechi, R.\ Giachetti, E.\ Sorace, M.\ Tarlini,
{\it Deformation Quantization of the Heisenberg Group},
Commun.\ Math.\ Phys. {\bf 169}  (1995), 627.

\bibitem{cuza} T. L. Curtright, G. I. Ghandour, C. K. Zachos, J. Math. Phys.
{\bf 32} (1991), 676. 

\bibitem{Sweedler}
M.\ E.\ Sweedler, {\it Hopf Algebras}, (Benjamin, New York, 1969).

\bibitem{SWZ}
P.\ Schupp, P.\ Watts, and B.\ Zumino,
{\it Bicovariant Quantum Algebras and Quantum Lie Algebras},
Commun. Math. Phys. {\bf 157} (1993), 305.

\end{thebibliography}
\end{document}